# White Organic Light-Emitting Diodes with fine chromaticity tuning *via* ultrathin layer position shifting

Hakim Choukri, Alexis Fischer, Sébastien Forget[*], Sébastien Chénais, M-C Castex
*Laboratoire de Physique des Lasers, Université Paris 13 / CNRS, 93430 Villetaneuse, France*

Dominique Adès, Alain Siove
*Laboratoire de Biomatériaux et Polymères de Spécialité, Université Paris 13 / CNRS, 93430 Villetaneuse, France*

Bernard Geffroy
*Laboratoire Composants Hybrides, LITEN/DTNM/CEA-Saclay, 91191 Gif/Yvette, France*

**Abstract :** Non-doped white organic light-emitting diodes using an ultrathin yellow-emitting layer of rubrene (5,6,11,12-tetraphenylnaphtacene) inserted on either side of the interface of a hole-transporting α-NPB (4,4'-bis[*N*-(1-naphtyl)-*N*-phenylamino]biphenyl) layer and a blue-emitting DPVBi (4,4'-bis(2,2'-diphenylvinyl)-1,1'-biphenyl) layer are described. Both the thickness and the position of the rubrene layer allow fine chromaticity tuning from deep blue to pure yellow *via* a bright white (WOLED) with CIE coordinates (x= 0.33, y= 0.32), a $\eta_{ext}$ of 1.9%, and a color rendering index (CRI) of 70. Such a structure also provides an accurate sensing tool to measure the exciton diffusion length in both DPVBi and NPB (8.7 and 4.9 nm respectively).

Organic light emitting devices (OLEDs) are a promising technology for fabrication of full-color flat-panel displays. The development of OLEDs relies on the capability to obtain emission spanning the full visible spectrum. In particular, White OLEDs (WOLEDs) are of foremost interest for lighting and display applications[1]. To achieve white emission, various methods have been used, such as *e.g.* excimer/exciplex emission[2], mixing of three (red, blue, green) or two (complimentary) colors in a single host matrix — or in physically separate layers[3]. Among these various devices, numerous doped-type WOLEDs using two mixed complimentary colors to produce white have been fabricated[4,5]. Although the co-evaporation process allows to a certain extent a control of the emitted radiation color *via* the different evaporation rates, it remains technologically difficult to accurately control the concentration. Hence, fine tuning of the color and achievement of bright white emission remain problematic. We report in this letter on a way to finely tune the color, including balanced white emission, in a multilayer non-doped OLED[6,7] based on blue matrices, in which an ultrathin yellow emitting layer was inserted. We show that by adjusting both the thickness and position of this layer, a very accurate control of the emitted color can be obtained, from deep blue with CIE coordinates (0.17, 0.15) to pure yellow (0.51, 0.48), *via* a bright white (0.32, 0.31) close to the equi-energy white point (0.33, 0.33), and a quite good Color Rendering Index (CRI) of 70. The external quantum efficiencies, the chromaticity coordinates and the luminance values are investigated for various thicknesses and positions of the yellow-emitting layer. Finally, the device structure, sometimes referred as "delta doping"[8], allows a better understanding of the emission process through an experimental determination of the exciton lengths.

The 0.3 cm$^2$-active-surface OLED-structure consists of the different layers described in fig. 1. The Indium Tin-Oxide (ITO)-covered glass substrate was cleaned by sonication in a detergent solution, then in deionized water and prepared by a UV-ozone treatment. Organic compounds were deposited onto the ITO anode by sublimation under high vacuum (10$^{-7}$ Torr) at a rate of 0.1 – 0.2 nm/s. An *in-situ* quartz crystal was used to monitor the thickness of the vacuum depositions. We sequentially deposited onto the ITO anode a thin (10 nm) layer of copper pthalocyanine (CuPc) as a Hole-Injecting Layer, then a 50 nm thick α-NPB (4,4'-bis[*N*-(1-naphtyl)-*N*-phenylamino]biphenyl) layer as a Hole Transporting Layer (HTL). A 60 nm thick DPVBi (4,4'-bis(2,2'-diphenylvinyl)-1,1'-biphenyl) layer is then deposited and acts as a blue emitter matrix and Electron

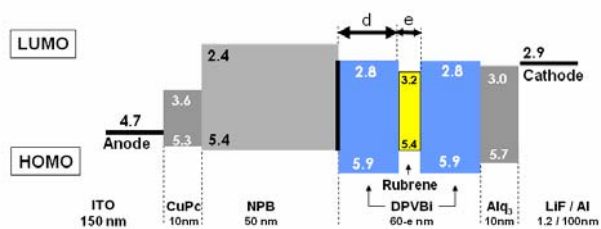

*Figure 1 : Structure of the OLEDs. Energy levels are taken from the literature. d can be either positive or negative.*

---

*\* e-mail : forget@galilee.univ-paris13.fr*

Transporting Layer (ETL). Finally, a thin (10 nm) Al(Q)$_3$ (Aluminium tris(8-hydroxyquinolinate)) layer was used as electron-Injecting layer onto the LiF/Al top cathode. An ultrathin layer of Rubrene (5, 6, 11, 12-tetraphenylnaphtacene) was inserted within the structure, at a distance $d$ from the NPB/DPVBi interface. The position $d$ of this layer was varied between -10 and +20 nm *i.e.* either within the NPB layer ($d<0$) or the DPVBi layer ($d>0$). The choice of rubrene is the consequence of its high photoluminescence quantum yield (near 100%)[9], as well as the very efficient Förster energy transfer from DPVBi due to the good overlap between the rubrene absorption and the DPVBi emission spectra (inset in Fig. 2). Electroluminescence spectra and chromaticity coordinates were recorded with a PR 650 SpectraScan spectrophotometer at a constant current density of 30mA/cm². All the measurements were performed at room temperature and under ambient atmosphere, without any encapsulation.

Assuming a vacuum-level alignment across the heterojunction, the NPB/DPVBi interface appears as a barrier for holes coming from the anode (ΔE ~ 0.5 eV between HOMO levels) as well as for electrons coming from the cathode (ΔE ~ 0.4 eV between the LUMO levels). Excitons are then expected to recombine mostly around this interface: similar structures have proved to be efficient for blue-light emission[6,10].

The DPVBi thickness has to be relatively thin, as the electron mobility in DPVBi[11] is several orders of magnitude lower than the hole mobility in NPB: the electrical performances consequently depends on the DPVBi thickness[12,13]. From the optical point of view, we also have to take into account the microcavity effects (between the glass/ITO interface and the top Al layer). Consequently, the DPVBi total thickness was optimized to 60 nm so that the recombination zone (*i.e.* the NPB/DPVBi interface) was located at the first antinode of the standing wave in the microcavity (for the peak wavelength of DPVBi $\lambda_{max}$= 456 nm). Since the ultrathin layer of rubrene is never more than 20 nm away from the interface, cavity effects play a negligible role on the color balance here.

Table 1 : Effect of position of the 1-nm wide rubrene layer : external quantum efficiency $\eta_{ext}$(%), power efficiency (lm/W), current efficiency (cd/A), luminance (cd/m²), and CIE coordinates (x,y) of the devices @ 30 mA/cm².

| d(nm) NPB | d(nm) DPVBi | $\eta_{ext}$(%) | (lm/W) | (cd/A) | (cd/m²) | CIE (x,y) |
|---|---|---|---|---|---|---|
| -1 | _ | 1.2 | 1.1 | 4.4 | 1584 | (0.51, 0.48) |
| -3 | _ | 1.4 | 0.9 | 4.1 | 1689 | (0.39, 0.38) |
| -3.5 | _ | 1.7 | 1.1 | 4.1 | 1795 | (0.32, 0.31) |
| -5 | _ | 1.7 | 0.9 | 3.5 | 1700 | (0.24, 0.27) |
| -10 | _ | 3.4 | 1.2 | 4.4 | 2275 | (0.17, 0.15) |
| _ | 0 | 1.2 | 1.2 | 4.6 | 1600 | (0.51, 0.48) |
| _ | +5 | 2.6 | 2.5 | 8.2 | 4067 | (0.41, 0.43) |
| _ | +10 | 2.0 | 1.0 | 3.7 | 1700 | (0.25, 0.23) |
| _ | +20 | 2.8 | 1.1 | 3.6 | 1900 | (0.17, 0.15) |

In order to investigate the color-control potential and the performance of our structure, we first varied the rubrene layer thickness (e) on either side of the NPB/DPVBi interface (while keeping the total OLED thickness constant). The external quantum efficiency $\eta_{ext}$ (%) with increasing rubrene thickness is given in Fig. 2 (on the DPVBi side only for clarity). We observe a decrease of $\eta_{ext}$ when the Rubrene thickness varies from 1 to 10 nm: this results have already been observed with other fluorescent dyes and could be attributed to fluorescence quenching[14].

Moreover, the electrons coming from the cathode could be trapped in the Rubrene layer due to the high difference of LUMO between DPVBi and Rubrene (0.4 eV), and the turn-on voltage $V_{th}$ is consequently increased from 5.5 to 7V when (e) is increased from 0 (no rubrene) to 10 nm. We also observe the same decrease on the NPB side, although less abrupt as the trapping effect is much less sensitive in this case owing to the quasi-alignment of HOMO levels (the change in $V_{th}$ in less than 0.5V). The optimum thickness is 1 nm, which has to be related to the size of a rubrene molecule (~1.7 x 1.4 x 0.7 nm), so that the ultrathin layer may be thought as an intermediate case between the monolayer and a "localized doping" situation. The carrier transport is not affected by such a thin layer, as confirmed by the constant value of $V_{th}$ with or without it. Indeed, the carriers are expected to tunnel easily over this quasi-monolayer of rubrene without being trapped. Moreover, the much localized yellow emitter offers an efficient probe for measuring exciton diffusion length (see later in the text). In the following the rubrene thickness is kept equal to 1 nm.

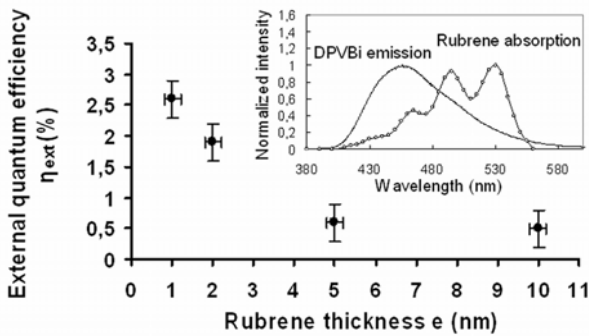

*Figure 2 : External quantum efficiency $\eta_{ext}$ (%) versus Rubrene thickness in DPVBi (d = 5 nm) at current density 30 mA/cm².*
*Inset : rubrene absorption and DPVBi emission spectra.*

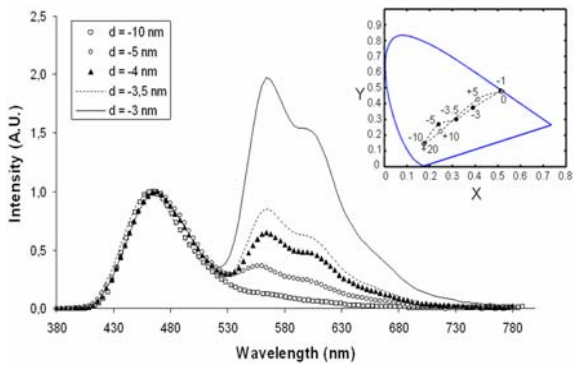

*Figure 3 : EL spectra for different positions of rubrene in the NPB (the DPVBi side is similar but was not plotted for clarity). Inset : Corresponding CIE diagram : filled circle stand for DPVBi side (the values of d, here positive, is written near each point) and open circles stand for NPB side. The dotted line is a guide to the eye.*

The influence of the position $d$ of the rubrene layer is summarized in Table 1. Typical emission spectra for different values of $d$ are plotted on fig.3. In these configurations, the emission spectrum consists in a blue and a yellow peak. The relative intensity of these peaks is well controlled by the position of the rubrene layer with respect to the interface, on both sides: when the rubrene layer is far away from the NPB/DPVBi interface (on either side), the emitted spectrum is dominated by the DPVBi emission, leading to a deep blue OLED with CIE coordinates (0.17, 0.15) corresponding to the DPVBi photoluminescence spectrum and a good quantum efficiency of 3.4%. Pure yellow emission with CIE coordinates (0.51, 0.48) corresponding to the emission light of rubrene molecules is achieved when the rubrene lies exactly at the interface ($d$ = 0). Here both Förster energy transfer as well as direct exciton trapping in the rubrene layer are very efficient and allow explaining the observed spectrum. Finally, white emission with CIE coordinates (0.32, 0.31), very close to the equi-energy white point (0.33, 0.33) was obtained for a position $d$ = -3.5 nm inside NPB. Noteworthy, a similar white light (0.33, 0.32) was obtained with a slightly thicker layer (2 nm) located inside DPVBi at $d$ = +5 nm (with a luminance of 2234 cd/m²). The quantum efficiencies are respectively 1.7% and 1.9% for the two white devices (e = 1 or 2 nm): those values are lower than the blue OLED, showing that some quenching still limits the efficiency. The CRI of the white devices have been calculated to be around 70. The CIE coordinates for different positions of the 1 nm rubrene ultrathin layer are plotted in the inset of Fig. 3. The emitted color of the device can be continuously adjusted from deep blue to pure yellow, along the line joining the CIE coordinates of DPVBi and of rubrene emission, with a simple shift of the rubrene layer position. This level of control cannot be easily achieved with the classical co-evaporation process where the yellow emitter is included, more or less homogeneously, in the whole blue layer. We also avoid the spectral shift often observed when the fluorescent molecule concentration increases, a phenomenon mainly attributed to polarization effects[15].

Another interesting observation in our devices is the symmetry of the system: color mixing leads to white-emitting devices whatever the rubrene layer is inside NPB or DPVBi. This is very different from what is observed in the classical NPB/Al(Q)$_3$ diodes, where the excitons only lie on the ETL side (Al(Q)$_3$)[16,17]. It has been demonstrated that in a rubrene-doped HTL/Al(Q)$_3$ device, the main emission mechanism is direct carrier recombination in rubrene[16]. In our non-doped devices, assuming vacuum-level alignment at the interface[18] (that is, neglecting charge-transfer induced dipoles), the electrons can hardly pass the NPB/DPVBi barrier, so that direct charge carrier trapping in the HTL is likely to be negligible. It appears that after the occurrence of the recombination process within a narrow zone around the NPB/DPVBi interface, excitons diffuse in both directions towards the anode and the cathode, eventually attaining rubrene either by direct capture or by Förster energy transfer, provided that the rubrene layer is not too far away. Our experiment offers a simple way to investigate exciton diffusion, by considering the rubrene layer as a small scale probe (sensing layer), as already demonstrated in so-called "delta-doping" experiments[6,8]. In fig. 4 we plotted the ratio of the rubrene emission peak intensity over the blue intensity as a function of $d$, which is well fitted by an exponential decay on both sides, as would be expected from a pure exciton diffusion mechanism[19,20]. The characteristic length of the decay $L_D$ (4.9 ±1nm in NPB and 8.7 ±0.6 nm in DPVBI matrix) is not strictly speaking the pure exciton diffusion length, but also contains the holes and electrons (if any) penetration depth as well as the Förster radius (typically around 3 nm). This supports the fact emission is controlled by the diffusion of excitons and their lifetime within the crossed layer *i.e.* NPB or DPVBi. This makes sense with the NPB and DPVBi having very similar energy gaps so there is no preferential diffusion direction for the excitons, contrary to the HTL/Al(Q)$_3$ case.

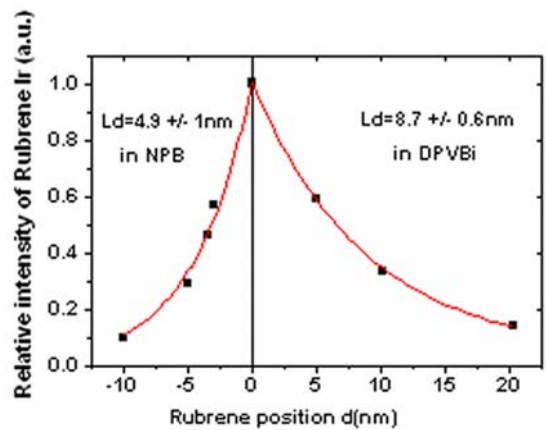

*Figure 4 : Relative Rubrene intensity for different positions in NPB and DPVBi. The dots show the experimental data, and the line fits the exponential*

In summary, we showed that by adjusting both the thickness and position of an ultrathin layer of rubrene on either side of the NPB/DPVBi interface, a very accurate control of the emitted color can be obtained. Starting from deep blue with CIE coordinates (x= 0.17, y= 0.15) to pure yellow with CIE (0.51, 0.48) *via* a bright white with excellent CIE coordinates (x= 0.33, y= 0.32), $\eta_{ext}$=1.9%, a quite good Color Rendering Index (CRI) of 70 and a good luminance of 2234 cd/m² at 60mA/cm² were demonstrated. The opportunity of inserting a thin layer on both sides of the recombination zone opens the door towards more complex structures using 3 or more different emitters at various wavelengths, with adjustable positions and thicknesses. Such a configuration may improve OLED color control and particularly the CRI value.